# Platform Based Design for Automotive Sensor Conditioning

L. Fanucci[1], A. Giambastiani[2], F. Iozzi[3], C. Marino[3], A. Rocchi[2]

[1] IEIIT, National Research Council, Via Caruso, I-56122 Pisa, Italy
Tel: +39 050 2217.668. Fax: +39 050 2217.522, e-mail: l.fanucci@iet.unipi.it
[2] SensorDynamics AG, Via Giuntini 25, I-56123 Navacchio (Pisa), Italy
[3] Dept. of Information Engineering, University of Pisa, Via Caruso, I-56122 Pisa, Italy

**Abstract**

*In this paper a general architecture suitable to interface several kinds of sensors for automotive applications is presented. A platform based design approach is pursued to improve system performance while minimizing time-to-market.. The platform is composed by an analog front-end and a digital section. The latter is based on a microcontroller core (8051 IP by Oregano) plus a set of dedicated hardware dedicated to the complex signal processing required for sensor conditioning. The microcontroller handles also the communication with external devices (as a PC) for data output and fast prototyping. A case study is presented concerning the conditioning of a Gyro yaw rate sensor for automotive applications. Measured performance results outperform current state-of-the-art commercial devices.*

## 1. Introduction

The increase of reliability and performance brought by electronic circuits in automotive industry is leading to the progressive replacement of predecessor mechanical and electromagnetic devices. Furthermore, electronics made possible several new applications in car's safety (e.g. airbag, ABS, TCS, ESP, proximity sensor) and/or comfort (e.g. driving assistance, air conditioning): all these advanced facilities are based on a continuously growing set of sensors, spread all over the car. As they get more sophisticated, according to market trends and requirements, they also need more accurate and reliable sensing elements; as well the high number of sensors (more than 100 at present time) will inevitably drive future choices on the cheaper and less-consuming ones.
This rapid growth has recently lead to the necessity of developing new design methodologies, allowing the designers to keep up with the increasing complexity of design and decreasing time-to-market requirements [1]. In order to deal with these issues, several solutions have been proposed, among which the so called Generic (or Universal) Sensor Interface [2-5]. This kind of interface approach integrates on chip all the analog/digital resources that are needed to perform conditioning for a wide range of sensors (capacitive, resistive, inductive, etc.). A cost and time-to-market reduction is apparently achieved, as one chip can interface several sensors (often simultaneously), providing only the essential circuitry, while non-standard customized functions are left to an external microcontroller. Indeed, these sub-optimal architectures lead, for a given sensor, to an increase in overall area and power consumption, and a reduction of overall performance and safety level (automotive sensors can be very different from each other), if compared to chips dedicated to a specific class of sensors.

In this paper a platform-based methodology for a generic automotive sensor interface is presented, in order to overcome the Universal Sensor Interface limitations while reducing time-to-market. Platform-based approach defines the electronic system design as sequence of several abstraction layers (each one can be considered as a platform) [6]. Concerning an embedded system, a platform can be defined as a set of modules, interfaces, services and software that should be as much as possible configurable. They are built up taking into account the wide-ranging signal conditioning electronics for different sensors of modern automobiles, in a way that from such generic platform, the optimum interface for a specific sensor can be easily derived in a short time by means of system simulation, verification and possibly prototyping. This entails that only the required analog/digital components are integrated onto silicon, resulting in minimized area and power overheads.

## 2. Platform based design flow

Our approach is initially based on the realization of a MATLAB™ model for the system at the highest abstraction level, which is made of a set of functional blocks with no distinction between analog/digital sections and software. The sensor itself can be modeled with MATLAB™, and thus co-simulated with the conditioning circuitry, helping the designer to find the most appropriate elaboration chain for the given sensor. Through simulations, design iterations and functional blocks refinements a project space exploration can be performed, together with the definition of more detailed descriptions of each block's required functionalities. Such investigation allows a first partitioning of the system in analog, hardwired and programmable (software) digital building blocks: even if this subdivision cannot be taken as frozen, the accurate MATLAB™ modeling and simulations guarantee the reasonability of these choices.

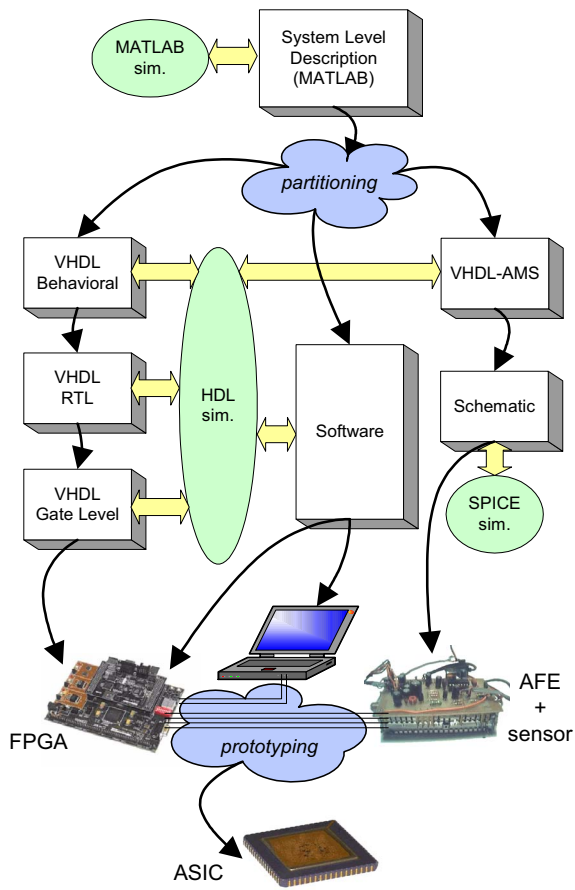

**Fig. 1. Design flow**

Then each block is modeled with the most appropriate description language: VHDL for digital hardware, VHDL-AMS for analog circuitry and C/C++ for software routines. The top-down design flow is depicted in Fig. 1: from the initial behavioral model we get to lower levels of abstraction via synthesis steps. The result of a synthesis step is then validated with the previous one through a verification phase. For what concerns the digital section, through a RTL (Register Transfer Level) description and a synthesis tool we produce a Gate-Level VHDL for the selected technology.

Software development goes along with digital Intellectual Property (IP) macrocells progress: low level drivers are provided just after the first stable VHDL (behavioral), so that general high level software can be developed almost independently from the IP implementation

On the analog side, from VHDL-AMS (modeling no more than the specifications) we generate a transistor-level description, from which a more accurate VHDL-AMS model can be obtained and employed in a mixed simulation, together with standard VHDL and C software running on the programmable hardware (microcontroller or general purpose processor). After the implementation of each single block, the relevant behavioral model is updated, so that it behaves closer to the practical circuit. This feedback process increases mixed simulations efficiency, allowing a more comprehensive design space exploration and reducing the probability of system architecture re-design [1]. VHDL-AMS becomes therefore crucial to let designers simulate the whole system during its own development, compare results with those given by MATLAB™ and thus be able to fix as soon as possible unexpected behaviors, by trimming the architecture and updating blocks functionalities or specifications.

The above mentioned design flow ends up with the prototyping phase, through which we can quickly test the whole system (or parts of it) in real operating conditions. Analog hardware can be implemented at first with discrete components mounted on a board (as our aim is to keep the analog section as simple as possible, this emulation should not differ too much from the final ASIC implementation, giving reliable results). As well it can be later mapped onto a dedicated analog chip, which can be used as a reference for digital hardware and software prototyping. The digital section is tested by means of a proper board, whose core is a large FPGA, provided with connectors to link an external PC and the analog front end. This system prototype allows the designers to verify the correct interaction between analog and digital parts, to develop firmware on a real platform implementation (out of the slow simulation environment) and, above all, to have a first, reliable estimate of system performance, from which architectural improvements can be further derived. Such prototyping phase is essential to get to the ASIC implementation with good confidence and high chance of success at first attempt: FPGA and analog front end not only have to satisfy functional specification for the targeted sensor, but also have to pass strict self-checking tests concerning full hardware read-back capability.

## 3. Generic platform for sensor interface

The generic platform shown in Fig. 2 is intended to address the design of the sensor interface for a wide range of automotive applications. It is based on three main parts: analog, hardwired digital and programmable digital (software).

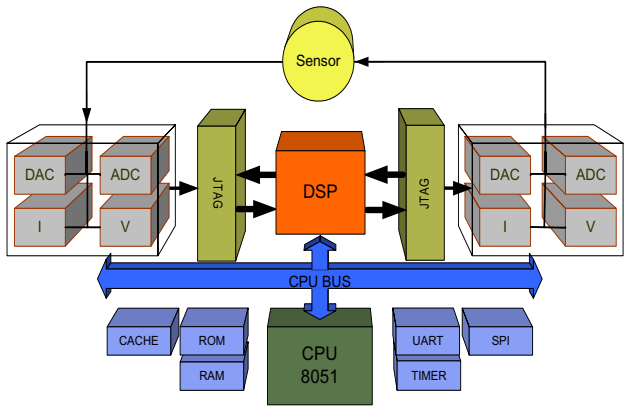

**Fig. 2. Platform architecture**

For the analog section, it can be critical in automotive applications to meet the required noise, area and power consumption constraints within a wide temperature range (-40°C÷125°C). This is particularly true if compared with the digital counterpart: for this reason we try to use as little analog signal processing as possible and to perform most of the conditioning in the digital domain. Thus the analog front-end only consists of ADCs, DACs, amplifiers and voltage/current sources, which are essential building blocks for automotive sensors conditioning. All the other platform functions and services are implemented either through digital hardware or software, keeping as much as possible fixed, simple and thus reliable the analog section. The front-end can be customized for different classes of sensors (inductive, capacitive, etc.) by choosing the most suitable analog cells from a well-stocked IP portfolio. Moreover programming main components parameters (such as amplifier gains and bandwidth, number of ADC bits, etc.) through the digital part allows a more accurate adaptation of the front end circuitry to the requirements of different sensors, both at design stage and during real working conditions (with the chance of on-line trimming of such parameters).

All non-trivial signal processing required for sensor conditioning is performed by the digital section, which also monitors system activity and handles communication with external devices (such as a PC during the prototyping phase). The monitoring/control functions are particularly useful for platform development and customization on the given sensor. Two kinds of computing resources are available: dedicated IPs for digital signal processing (DSP block in Fig. 2), such as FIR/IIR filters, modulator, demodulator, etc., and a general purpose processor or a microcontroller (an 8051 CPU core has proved to be sufficient up to the current development) provided with memories, buses and peripherals for communication. The CPU is mainly in charge of monitoring the signal processing chain and providing output status data through standard interfaces such as UART and SPI. This partitioning between processing and monitoring functionalities is important to have maximum performances without hardware overhead (only the required blocks for both processing and monitoring will be finally implemented on silicon) and at the same time guarantee fast and accurate platform optimization for the given sensor.

## 4. A case study

In this section we present a case study for the above mentioned platform based design flow, consisting in a customization of the architecture described in section 3 for a gyroscopic angular rate sensor, as implemented by SensorDynamics AG for a commercial chip.

### 4.1 Gyro sensor

Vibrating ring gyroscopes consist of a circular ring provided with drive, sense and control electrodes [7]. Even though they are also available as discrete components, we focused on MEM implementation [8], being our aim to carry out a stand-alone device able to provide accurate yaw rate measurement.

Gyro sensors base their working on the Coriolis force acting on a vibrating mass: while the driving electrodes keep the ring vibrating along the primary direction (with fixed amplitude, around the z-axis in Fig. 3 [8]), the rotation of the device ($\Omega_x$) causes an energy transfer to the second vibrating mode, which is located at 90° from the primary mode (on the y-axis). The amplitude of this vibration is proportional to the angular rate, and can be capacitively detected through the sense electrodes placed on the rectangular structure (open loop mode). A closed loop configuration exploits the control electrodes, by means of which the secondary vibration can be compensated, in order to let the sensor work around its rest point, thus achieving more linear and accurate measures.

Such sensors basically require a PLL (for primary drive), which has to keep the ring in resonance (at a frequency of approximately 15 KHz), an AGC (to control the amplitude of this vibration) and a chain including demodulators, filters, temperature/offset compensation and modulators for secondary drive and rate sensing.

### 4.2 System architecture

The generic platform architecture depicted in Fig. 2 has been customized to the gyro sensor conditioning requirements, in order to achieve an almost optimal implementation for interfacing this kind of sensor, together with several communication resources for easier debugging, development and performance improvement.

For what concerns the analog front end, it only absolves functions of driving sensor's electrodes (through couples of DACs for each loop) and performing signal acquisition (by means of SAR ADCs, amplifiers and basic filters). It also provides stable power supply and clock to the digital section.

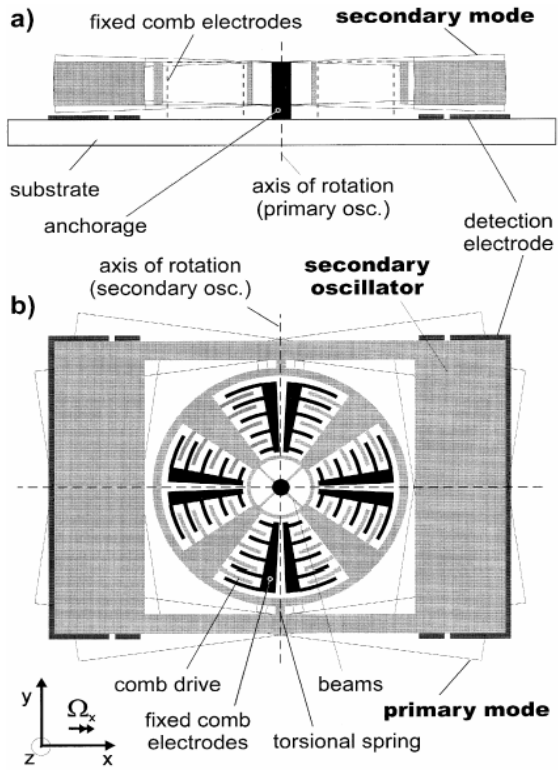

**Fig. 3. Gyro sensor.
(a) Cross-section.   (b) Top view.**

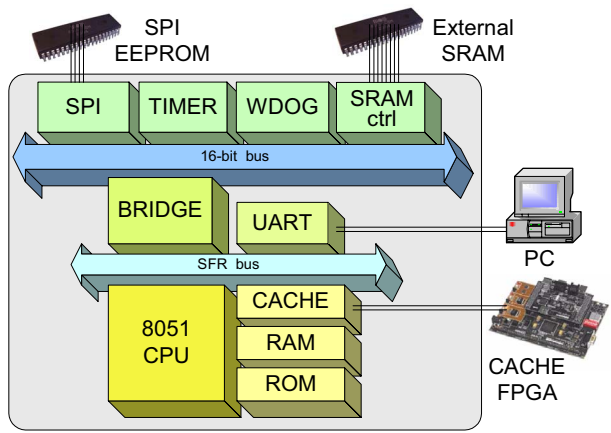

**Fig. 4. CPU core architecture with communication resources.**

Each analog cell in the front end is digitally controlled, and this programmability can be of paramount importance for the whole system functioning. Particular effort has been placed in building a reliable and efficient interface between the analog and digital world. A JTAG standard interface has been selected for the following main reasons: (i) the standard protocol has largely been studied and tested for several applications, so it guarantees high reliability since first implementation; (ii) JTAG bases on asynchronous communication, which limits clock skew issues that may lead to undetectable faults; (iii) it employs a short number of wires (only 4 per chain), thus resulting easy to route also on very complex chips; (iv) it allows for full read-back capability (for fast verification or debugging).

The digital section performs almost all the signal processing (within the DSP block), and manages the communication resources (through the CPU core). The DSP contains a chain of IPs for signal elaboration, featuring the blocks schematically depicted in Fig. 4.: its architecture and sub-blocks dimensioning are derived from the MATLAB™ model set up during the design space exploration phase. Though all the required processing can be performed independently by analog front end and DSP block, a CPU core is also present to fulfil control, monitoring and communication tasks. These functionalities naturally find their best implementation via software routines, as they may vary through system updates and new system requirements.

Control and monitoring are performed real-time by the processor on both DSP and analog front end: a routine constantly checks the system status by accessing the several readable registers spread along the processing chain (for example makes sure that the PLL is locked). Meanwhile other routines handle communication services, providing status and output data to the user: during prototyping phase, the system can be linked to a PC and through a graphical interface manual trimming can be performed and all intermediate data of the chain can be accessed. The same communication resources are used during the chip normal working conditions to provide external devices (typically the car's electronic control system) the required angular rate measure and status information.

As shown in Fig. 4, CPU core architecture comprises the Oregano 8051 processor [9] (freely distributed under LGPL licence), which provides a good compromise between performance and area occupation, and fits well the mentioned microcontroller applications; it is provided with ROM/RAM memories and cache controller, all configurable (both with hardware generics and at run-time) in order to get the maximum flexibility for software download, development and update. Just to give a few examples, an 'ASIC' version could include a big ROM (16 Kb) with all the needed software (the latest available at the moment) and through the cache (which is conceived to access big external RAM with a custom 2-wire protocol) newer software versions could be downloaded and tested; in a 'prototype' version, a big RAM would be instantiated and used as Program Storage (while the boot placed in a small 1 Kb ROM would perform software download via UART) and cache would not be instantiated. Software download is also possible by means of RS485 (in place of simple RS232 protocol implemented by the UART) and SPI (at start-up all the communication devices look for a response on their channel, in a way that the connected peripheral is automatically detected); moreover it's possible to store the downloaded software into an external SPI EEPROM, and so reboot directly from EEPROM instead of downloading each time after reset. This high configurability gives the designers the chance of developing software with maximum simplicity and efficiency, through the whole system prototyping phase and even after the first ASIC releases, in order to achieve the utmost confidence on the final product success. Cache controller and UART are located on the 8051 Special Function Register (SFR) Bus (8-bit), while the other peripherals (SPI, timer, watchdog, and SRAM controller) are accessed via a custom bridge by means of a 16-bit bus. SRAM controller is used during the prototyping phase, to store at real-time (into a 512 Kb SRAM) digital data coming from any node of the DSP chain, with chance of later read-back for analysis purposes.

### 4.3 Results

The platform based design flow has been carried out step by step as described in section 2: simulation of the entire system (in particular sensor locking) has been first performed with MATLAB™ model (see Fig. 5 for main PLL signals), and then the same result has been reached in a HDL simulation environment, thanks to the full VHDL-AMS modelling of the analog section and the sensor itself. After this important achievement, designers have been able to get to the final mixed-signal platform implementation in a short time, being already defined block partitioning, dimensioning and connections.

The prototyping phase has then proved the validity of this approach, and an emulation environment has brought real sensors to locking (see Fig. 6 for measured PLL data) and output yaw rate data. The digital part of roughly 200 Kgates complexity has been implemented in a Xilinx X2S600E running a 20 MHz clock frequency.

Performance of higher level (as reported in Table 1, in comparison with other commercial gyro sensors of tables 2 and 3) have been achieved with the integration of the analog front end into a 12 mm$^2$ custom chip implemented in a 0.35 μm CMOS technology.

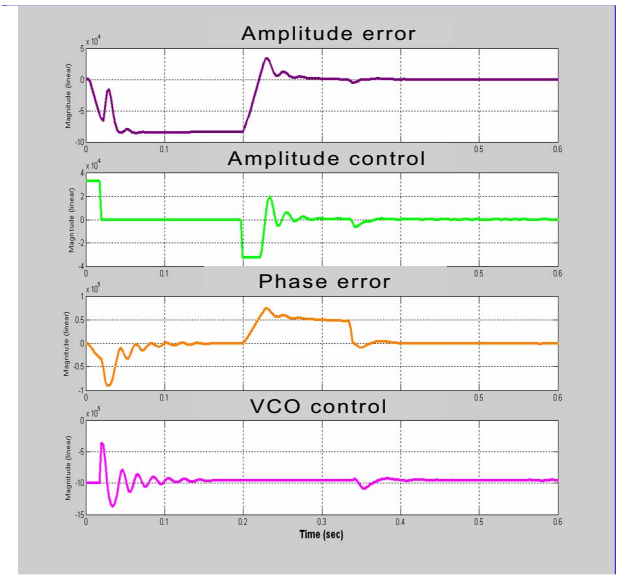

**Fig. 5. Waveforms of PLL locking (MATLAB).**

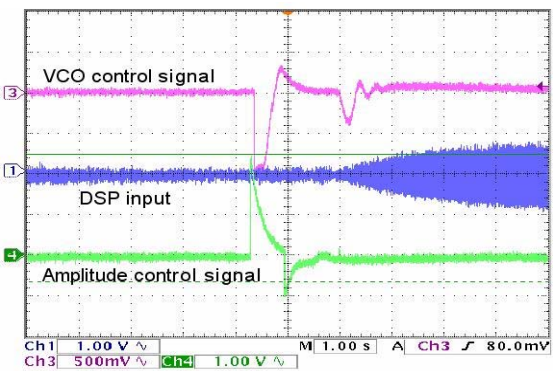

**Fig. 6. Measured waveforms (AC probe).**

| Parameter | SensorDynamics | | | Units |
|---|---|---|---|---|
| | Min. | Typ. | Max. | |
| **Sensitivity** | | | | |
| Dynamic Range | +/- 75 | | +/- 300 | °/s |
| Initial | 4.85 | 5.00 | 5.15 | mV/°/s |
| Over Temperature | 4.80 | 5.00 | 5.20 | mV/°/s |
| Non Linearity | 0.07 | 0.10 | 0.20 | % of FS |
| **Null** | | | | |
| Initial | 2.70 | 2.50 | 2.53 | V |
| Over Temperature | 2.70 | | 2.53 | V |
| Turn On Time | | 500.00 | | ms |
| **Noise** | | | | |
| Rate Noise Dens. | 0.04 | 0.09 | 0.13 | °/s / √Hz |
| **Freq. Response** | | | | |
| 3 dB Bandwidth | 25.00 | | 75.00 | Hz |
| **Temp. Ranges** | | | | |
| Operating Temp. | - 40 | | + 85 | °C |

**Table 1. Performance of SensorDynamics implementation.**

| Parameter | Analog Devices | | | Units |
|---|---|---|---|---|
| | Min. | Typ. | Max. | |
| **Sensitivity** | | | | |
| Dynamic Range | +/- 300 | | | °/s |
| Initial | 4.60 | 5.00 | 5.40 | mV/°/s |
| Over Temp. | 4.60 | 5.00 | 5.40 | mV/°/s |
| Non Linearity | | | 0.10 | % of FS |
| **Null** | | | | |
| Initial | 2.30 | 2.50 | 2.70 | V |
| Over Temp. | 2.30 | | 2.70 | V |
| Turn On Time | | 35.00 | | ms |
| **Noise** | | | | |
| Rate Noise Dens. | | 0.1 | | °/s / √Hz |
| **Freq. Response** | | | | |
| 3 dB Bandwidth | | 40.00 | | Hz |
| **Temp. Ranges** | | | | |
| Operating Temp. | - 40 | | + 85 | °C |

**Table 2. Performance of AD XRS300.**

| Parameter | Murata | | | Units |
|---|---|---|---|---|
| | Min. | Typ. | Max. | |
| **Sensitivity** | | | | |
| Dynamic Range | | | +/- 300 | °/s |
| Initial | | 0.67 | | mV/°/s |
| Over Temperature | 0.54 | | 0.80 | mV/°/s |
| Non Linearity | - 5.00 | | + 5.00 | % of FS |
| **Null** | | | | |
| Initial | | 1.35 | | V |
| Over Temp. | | - | | V |
| Turn On Time | | - | | ms |
| **Noise** | | | | |
| Rate Noise Dens. | | - | | °/s / √Hz |
| **Freq. Response** | | | | |
| 3 dB Bandwidth | | | < 50 | Hz |
| **Temp. Ranges** | | | | |
| Operating Temp. | - 5 | | + 75 | °C |

**Table 3. Performance of Murata's Gyrostar.**

## 5. Conclusions

In this paper a mixed signal platform based design methodology for automotive sensor interface has been presented. This approach aims to fast identify and verify suitable analog/digital interface architecture, for a given sensor, before integration onto silicon. An outline of such architecture has been obtained from the platform based methodology, and its customization for gyro sensors conditioning (implemented by SensorDynamics) has been described in detail. Results achieved by this implementation (outperforming competitors' chips, in spite of the incomplete development stage) prove the validity of the proposed approach, not only in terms of performance (practically no area overhead and best fit circuitry), but also related to short time-to-market, due to the high reconfigurability of the general platform architecture. As well power consumption and final product cost are likely to be positively affected by optimal hardware selection and high level on-chip integration. The whole system including the analog and digital circuits is being integrated into a single chip in a 0.35 μm CMOS technology.